\documentstyle[11pt,psfig]{article}
\begin{document}
\parskip 1ex

\title{ NOVEL RECONSTRUCTION MECHANISMS: A COMPARISON BETWEEN
  GROUP-III-NITRIDES AND ''TRADITIONAL'' III-V-SEMICONDUCTORS}

\author{T. K. Zywietz, J. Neugebauer, M. Scheffler\\
Fritz-Haber-Institut der Max-Planck-Gesellschaft, Faradayweg 4--6,
  D-14195 Berlin, Germany\\
J. E. Northrup\\
Xerox Palo Alto
  Research Center, 3333 Coyote Hill Road, Palo Alto, California 94304}

\maketitle

\begin{abstract}
We have studied the driving forces governing reconstructions 
on polar GaN surfaces employing first-principles total-energy calculations. 
Our results reveal properties not observed for other semiconductors, 
as for example a strong tendency to stabilize Ga-rich surfaces. 
This mechanism is shown to have important consequences on various 
surface properties: Novel and hitherto unexpected structures 
are stable, surfaces may become metallic although GaN is a wide-bandgap 
semiconductor, and the surface energy is significantly higher than for 
other semiconductors. We explain these features in terms of the small 
lattice constant of GaN and the unique bond strength of N$_{2}$ molecules. 
\end{abstract}

\section{Introduction}
Progress in the materials quality of GaN has led recently to the first commercially
available, highly efficient optoelectronic devices emitting in the green and
blue/UV region \cite{nakamura94,nakamura97}. A substantial problem in growing
GaN and its alloys is the lack of a lattice-matched substrate. Bulk GaN can be
grown only in small crystallites and sapphire, the most commonly used
substrate, has an extremely large lattice mismatch (14\%). Another problem is
the high nitrogen vapor pressure of bulk GaN requiring highly activated
nitrogen precursors for the growth. Both issues make it difficult to grow
routinely and in a controlled fashion high quality GaN.  In order to overcome
these problems it is critical to understand the fundamental growth aspects on
an atomic level.

Initially, it has been believed that the driving forces behind surface
reconstructions and growth are essentially the same as for conventional
semiconductors such as {\em e.g.} GaAs. However, it soon became obvious that
GaN behaves in many aspects very differently. For example, impurity
concentrations are significantly higher than in conventional semiconductors
and p-type doping is rather difficult \cite{ponce}. Also, growth is much more
affected by even small changes in the growth environment and the structure of
the initial nucleation layer at the substrate controls the properties and
quality of the entire epitaxial layer.

Recent investigations also revealed surface structures for the technologically
relevant polar GaN surfaces which are very different from the well-established
structures of III-V semiconductor surfaces. Furthermore, some of these
structures disobey well-accepted empirical rules and models, which have been
found useful for understanding why and how semiconductor surfaces reconstruct.
For example, on cubic GaN (001) (which is commonly used to grow the cubic
phase of GaN) first-principles calculations identified a Ga-terminated surface
as the energetically most stable structure. On this surface four Ga atoms form
linear tetramers \cite{neugebauer98}. This is in contrast to conventional
semiconductors where dimers are the preferred building blocks on (001)
surfaces. Another example is the Ga adlayer structure which has been found
combining detailed STM and LEED measurements with first-principles
calculations at the wurtzite GaN $(000\bar1)$ surface \cite{smith}. This
surface violates several rules: It disobeys electron counting, atoms in the
top surface layer sit on singly coordinated sites, and each surface atom has
the highest possible number of dangling-bond states.

The aim of the present paper is to identify the properties of GaN that give rise to
these unusual surface reconstructions, to understand why the empirical rules that have been
well established to describe conventional semiconductor surfaces fail for GaN, and to determine
whether these rules can be extended to GaN. We will focus here on the {\em
  mechanisms} and {\em general principles} of the surface reconstructions.
Detailed descriptions of the specific atomic structures and the calculations for
cubic and wurtzite GaN can be found in Refs. \cite{neugebauer98,smith}. After a brief description of
the computational details (Sec. II) we will analyze the surface energies
of unreconstructed GaN and GaAs surfaces. Based on these results we show that
a characteristic feature of GaN surfaces is the tendency to have Ga atoms
in the surface layer. This feature, combined with the small lattice constant,
is shown to be responsible for the unusual surface reconstructions. Finally,
based on this analysis we derive conclusions concerning possible surface
reconstructions.

\section{Computational method}
The energy necessary to create a surface is called the surface energy. This
energy is not constant but depends on the specific thermodynamic conditions.
Specifically, in GaN the relative concentration of Ga and N atoms at the
surface determines the surface energy. The atomic reservoirs with which Ga and N
atoms are exchanged in order to modify the surface stoichiometry
determines the chemical potentials ($\mu_{\rm
  Ga}$, $\mu_{\rm N}$). The chemical potentials for Ga and N are not
independent variables, since for thermal equilibrium situations both species
are in equilibrium with the GaN bulk:
\begin{equation}
\mu_{\rm GaN}=\mu_{\rm Ga}+\mu_{\rm N}
\end{equation} 
The surface energy at $p = 0$ and $T = 0$ is then given by:
\begin{equation}
  \gamma=E^{\rm tot}-\mu_{\rm Ga}N_{\rm
    Ga}-\mu_{\rm N}N_{\rm N}
\end{equation} where $N_{\rm Ga}$ and $N_{\rm N}$ are the number of Ga and N
atoms and $E^{\rm tot}$ is the total energy of the surface obtained from
density-functional theory.  

The gallium chemical potential can be varied only between certain limits.
A major criterion is that the chemical potential for an
element is less than the chemical potential of the corresponding bulk material
(or molecules) since otherwise this element would form the energetically more
stable bulk or molecular structure. For the gallium chemical potential an
upper limit is therefore given if GaN is in thermodynamic equilibrium with
bulk Ga. This case is called the Ga-rich limit.  The lower limit is given for
GaN in thermodynamic equilibrium with N$_{2}$ molecules; it is therefore
called the N-rich limit. Using these relations and Eq. (1) we get:
\begin{equation}
  \mu_{\rm Ga(bulk)}+\Delta H_{\rm GaN} \le \mu_{\rm Ga} \le \mu_{\rm
    Ga(bulk)}\quad.
\end{equation}
Here, $\Delta H_{\rm GaN}$ is the heat of formation which is defined as:
\begin{equation}
  \Delta H_{\rm GaN}=\mu_{\rm GaN(bulk)}-\mu_{\rm Ga(bulk)}-\mu_{\rm
    N_{2}(molecule)}\quad.
\end{equation} A negative heat of formation means the reaction is exothermic.
The corresponding bulk chemical potentials are calculated from the bulk forms
of Ga metal (orthorhombic), N (N$_{2}$ molecule) and GaN (wurtzite). The total
energies have been calculated employing density-functional theory in the local
density approximation, in combination with a plane-wave basis set and
first-principles pseudopotentials. The exchange and correlation energy
functionals are those derived from the homogeneous electron gas calculations
of Ceperley and Alder \cite{ceperley}. We use soft Troullier-Martins
\cite{martins} pseudopotentials constructed with the fhi98PP
package \cite{fuchs98}. An explicit treatment of the Ga $3d$ electrons as
valence electrons has been found crucial to calculate accurate surface
energies. This required a large plane wave energy cutoff making our
calculations computationally rather challenging both with respect to CPU-time
and memory demand. We therefore used a parallel version of our plane wave code
on a Cray T3E. This version had been specifically optimized with respect to
data and CPU partitioning. Details about the program can be obtained from
Ref.  \cite{fhi96md}.

\section{Analysis of the surface structures}

As pointed out in Sec. 1, surface reconstructions of GaN exhibit features that
have not been observed on other III-V semiconductor surfaces. In order
to identify the mechanisms causing the unusual reconstruction we will analyze
the differences between polar GaN and GaAs surfaces. We will focus on polar
surfaces since non-polar surfaces show essentially the same features as found
for conventional semiconductors \cite{northrup96}.  Since we are here
interested in the mechanisms driving surface reconstructions on GaN surfaces
let us first briefly recall the rules and models that are typically applied
in a discussion of conventional semiconductor surfaces. A commonly used
principle is called the electron counting rule (ECR). According to the ECR the equilibrium
surface is one in which
the number of available electrons in the
surface layer will exactly fill all dangling-bond states in or near the
valence band and leave all states in or close to the conduction band empty
\cite{pashley}. An important consequence of this rule is that a surface
satisfying the ECR will be semiconducting. Further, just by counting electrons
and dangling-bonds a large number of potential surface structures can be
eliminated. While this rule is empirical, it has been found to work well for
almost all conventional semiconductor surfaces. Only a few exceptions have
been reported \cite{whitman,Biegelsen}.

The ECR is commonly justified in terms of Harrison's bond-orbital model
\cite{harrison}.  Atoms in conventional semiconductors are sp$^{3}$
hybridized. In the absence of reconstruction some of the hybrid orbitals cannot form bonds, but
instead give rise to partially occupied sp$^{3}$ dangling-bond states.
According to Harrison the energy levels of the cation dangling-bond states
are high in energy (lying close to or within the conduction band) and should
therefore be empty. Dangling bond states localized on the more electronegative
anions, however, are close to or within the valence band and should be
filled. Other mechanisms driving the reconstruction at
semiconductor surfaces are: (i) the tendency to reduce the number of dangling
bond states on the surface by forming e.g. dimers, adatoms or trimers
\cite{Biegelsen,Kaxiras} and (ii) minimizing the electrostatic energy by
optimizing the arrangement of charged surface atoms \cite{northrup94}.

\subsection{Reconstruction mechanisms for "traditional" semiconductors}

We will start our comparison between GaAs and GaN by considering the simplest
possible surface structures - the unreconstructed (1x1) surfaces. These
surfaces (Fig. 1) are terminated either by cations or by anions. For GaAs
possible structures are the open (001) surface and the close-packed (111)
(cation-face) and ($\bar1\bar1\bar1$) (anion-face) surfaces. Since the
equilibrium phase of bulk GaN is the wurtzite structure we performed the
calculations for (0001) (cation-face) and ($000\bar1$) (anion-face) which are
equivalent up to the fourth nearest neighbors to the cubic (111) and ($
\bar1\bar1\bar1$) surfaces. Since we are here only interested in qualitative
aspects, we will consider the (111)/($\bar1\bar1\bar1$) and
(0001)/(000$\bar1$) as equivalent surfaces. 

\subsection{Comparison between unreconstructed GaAs and GaN surfaces}

\setlength{\unitlength}{1mm}
\begin{figure}[tb]\centering
  \psfig{file=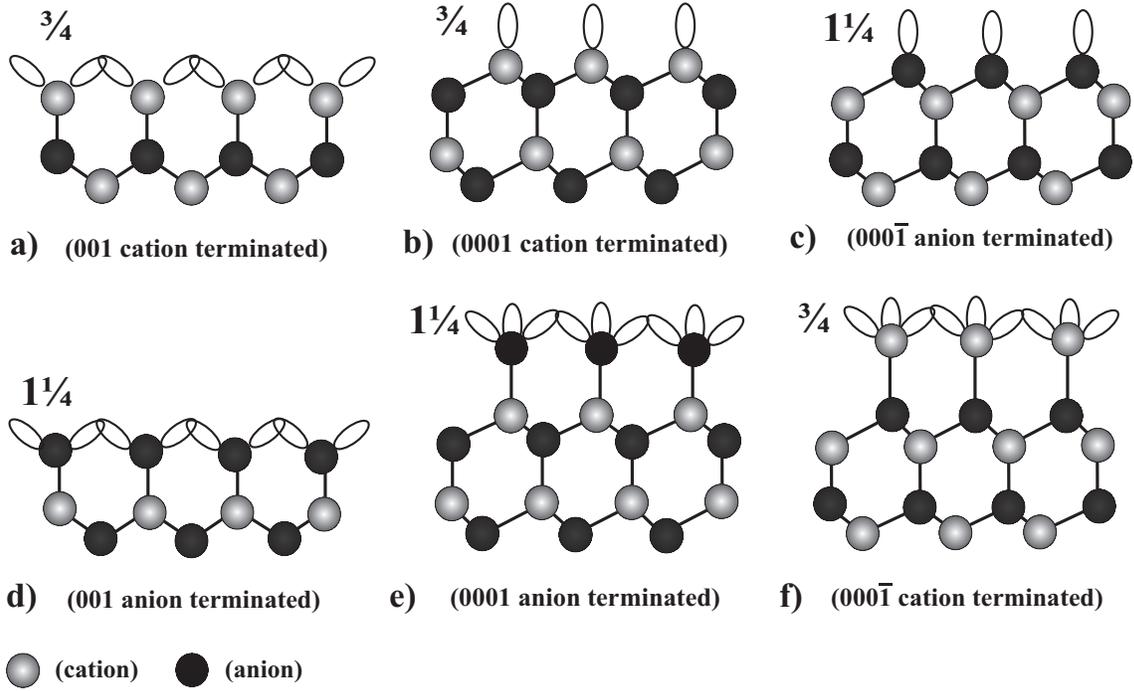,width=15cm}
\caption{Atomic structures for the low index (001), (0001) [which is
  equivalent to (111)] and (000$\bar1$) [which is equivalent to
  ($\bar1\bar1\bar1$)] surfaces of III-V-semiconductors. The numbers give the
  electrons per dangling-bond orbital.}
\end{figure}

The calculated surface energies of GaAs are plotted as a function of the chemical potential
in Fig. 2 (a-c). 
The energies are fully consistent with the
empirical rules and models discussed above. First, both Ga and As-terminated
(1x1) surfaces have much higher surface energies than the reconstructed
equilibrium surfaces implying that the (1x1) surfaces are unstable against
surface reconstruction. This is consistent with the fact that all possible
polar (1x1) surfaces have partially occupied dangling bonds and thus disobey
electron counting. Further, consistent with the principle of reducing the dangling
bond density, both the Ga-terminated (111) and the As-terminated ($\bar1\bar1\bar1$)
surfaces that have only {\em one} dangling bond orbital per surface atom (see Fig. 1b
and 1c) are energetically more stable than the corresponding As and
Ga-terminated surfaces that have {\em three} dangling-bond orbitals per surface
atom.

\setlength{\unitlength}{1mm}
\begin{figure}[tb]\centering
  \psfig{file=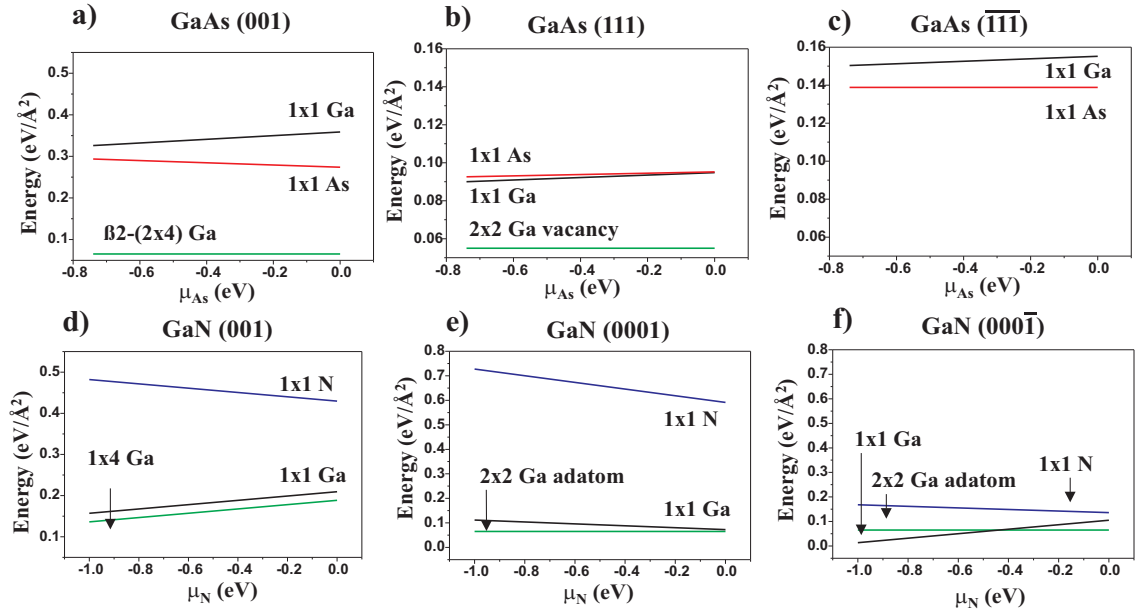,width=15cm}
\caption{Surface energies in eV/\AA$^{2}$ for the cubic (001), (111) and
  ($\bar1\bar1\bar1$) GaAs and the (001), (0001) and (000$\bar1$) GaN surfaces
  (solid lines). Note that both the unreconstructed Ga and As terminated
  surfaces are very high in energy compared to the equilibrium surfaces
  (dashed lines).}
\end{figure}

For GaN (Fig. 2, d-f) some of these mechanisms are no longer valid. First,
N-terminated surfaces are energetically always less stable than the
corresponding Ga-terminated surfaces. This result applies even for the
(000$\bar1$) surface where the Ga-terminated surface has three (Fig. 1f) and
the N-terminated surface only one dangling bond orbital per surface atom (Fig.
1c). Second, the Ga-terminated surfaces have a very low surface energy, which
is only slightly higher than the surface energy of the reconstructed
equilibrium surfaces. This is in clear contrast to GaAs where all
unreconstructed surfaces are much higher in energy than the equilibrium
structures. In fact, for (000$\bar1$) the Ga-terminated (1x1) structure (Fig.
1f) becomes the energetically preferred structure under Ga-rich conditions
(see Fig. 2f). This structure, however, obviously disobeys the electron
counting rule.

\subsection{Reconstruction mechanisms on GaN surfaces}

From these results we can immediately conclude that a major mechanism  driving
GaN surface reconstructions is the tendency to stabilize structures
that have more Ga than N atoms in the surface layer.  This conclusion is also
consistent with recent experimental and theoretical studies on GaN surface
reconstructions \cite{smith}.  Almost all equilibrium surfaces consist solely
of Ga atoms in the top surface layer. The only exception for polar surfaces
is the (0001) surface, where under nitrogen-rich conditions N-adatoms can be
stabilized on a Ga-terminated surface \cite{smith,bernholc97}.  This feature of
preferring just one species in the top surface layer (independent on the
chemical potentials) is unique to GaN and has not been reported for other
semiconductor surfaces. We will therefore attempt to elucidate the mechanisms
responsible for the stabilization of Ga atoms in the surface layer. According
to Eq. (2) we can seperate the surface energy into two contributions: (i) the
energy necessary to remove or add atoms to the chemical reservoirs (which
describe the specific growth conditions) and (ii) the total energy which is
the sum over all bond energies and includes contributions such as charge
transfer, electrostatic energy etc.

\subsection{Chemical Potentials}

Let us first focus on the chemical potentials. A lower limit on the energy
necessary to remove atoms from a chemical reservoir is the
bulk cohesive energy for solids and the binding energy per atom for molecules. These
energies are shown in Tab. 1 (taken from Ref. \cite{hocp} for a selected
number of elements). The elements have been chosen to be the constituent
species of the major semiconductor materials. Let us exclude for a moment
group IV elements which will be discussed at the end of Sec. 4.1. Among group
II, III, V, and VI elements nitrogen is obviously the element with the highest
binding or cohesive energy: The N-N bond in the N$_{2}$ molecule is one of the
strongest bonds found in nature. All other atoms have energies roughly between
2 - 3 eV, i.e. more than 2 eV less than a N atom in the N$_{2}$ molecule. We
can therefore conclude that for all compound semiconductors (except for group
III-Nitrides) there are only modest differences in the chemical potentials.
For group III-nitrides, however, there is a strong asymmetry in the chemical
potentials: More energy is required to transfer N atoms from the N reservoir
to the surface than to transfer Ga atoms to the surface.

\begin{table}\centering
\begin{tabular}{|c|c|c|c|c|c|c|c|c|c|c|}\hline
  Element & Zn & Al & Ga & In & N & P & As & O & S & Se \\ E$_{coh}$ (eV) &
  1.35 & 3.42 & 2.81 & 2.52 & 4.91 & 3.28 & 2.96 & 2.58 & 2.87 & 2.35\\
\hline
\end{tabular}
\caption{Experimental cohesive energies and molecule binding energies (N$_{2}$
  and O$_{2}$) of the constituent species in common semiconductors [15].}
\end{table}

A rough estimate of how this asymmetry affects the energy of the
GaN surfaces can be obtained by artificially eliminating the
large difference in the chemical reservoirs for Ga and N. We therefore shift
the Ga-chemical potential by the difference between the N and Ga chemical
potential (2.1 eV). The energy to remove a Ga atom is then the same as
removing a N atom from its chemical reservoir. The corresponding surface
energy is shown in Fig. 3 as dashed line. The energy significantly increases
(by $\Delta E_{\rm chem}$) bringing the surface energy closer to that of GaAs.

\setlength{\unitlength}{1mm}
\begin{figure}[tb]\centering
\begin{picture}(70,65)(0,-3)
  \psfig{file=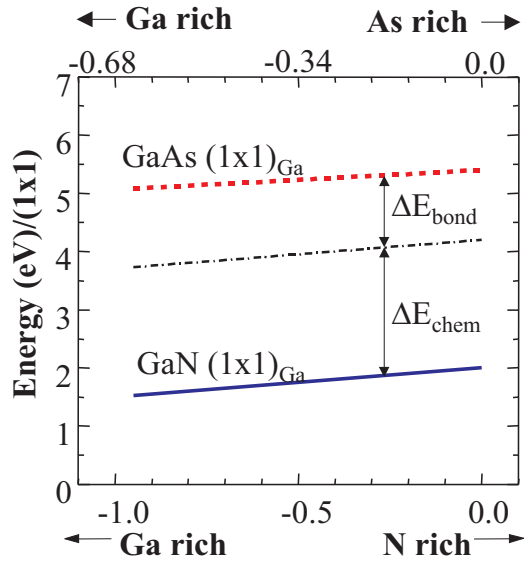,width=7cm}
\end{picture}
\caption{Surface energies for the unreconstructed Ga-terminated GaN and GaAs
  surfaces as a function of the Ga chemical potential. The dashed line
  corresponds to the Ga-terminated GaN surface where the difference between
  the Ga and N chemical potentials has been eliminated by artificially shifting
  the Ga chemical potential.}
\end{figure}

\subsection{Metallic bonding}

The fact, that $\Delta E_{\rm chem}$ gives only 2/3 of the difference with
respect to the GaAs surface energy (Fig. 3) indicates that the difference in
chemical potentials is not sufficient to explain the unusual stability of the
Ga-terminated surfaces at GaN. It is also determined by the binding energy an
atom gains if it is incorporated in the surface. The different binding
energies on both surfaces can be mainly understood by the formation of second
nearest neighbor bonds between surface Ga atoms. This effect can be roughly
estimated by calculating the formation energy of a free-standing Ga layer at
different lattice constants. The resulting energies shown in Fig. 4 reveal two
interesting aspects: First, the equilibrium lattice constant of the
free-standing Ga layer is close to the lattice constant of GaN bulk.
Therefore, the energy the Ga-layer gains when relaxing from the GaN-bulk
lattice constant to its ideal value is modest (0.2 eV). This feature implies
that at GaN surfaces the Ga atoms can form metallic bonds similar to those in
bulk Ga even without any relaxation.  Second, by going from the GaAs bulk
lattice constant to the GaN lattice constant the binding energy of the
free-standing Ga layer significantly decreases (by 0.9 eV). This energy gain
explains largely the stronger bonding energy ($\Delta E_{\rm bond}$ in Fig. 3)
of Ga atoms on the GaN surface. The energy reduction obtained by contracting the lattice
constant of a Ga adlayer from the GaN-bulk value to the equilibrium value
has been invoked to explain the stability of a laterally contracted incommensurate Ga-adlayer
structure on the GaN(0001) surface\cite{feenstra}.

\setlength{\unitlength}{1mm}
\begin{figure}[tb]\centering
\begin{picture}(70,65)(0,-3)
  \psfig{file=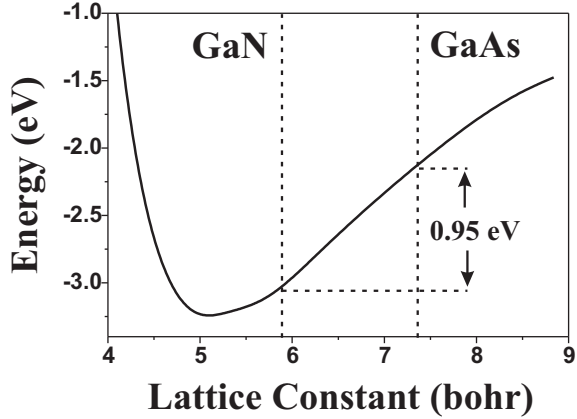,width=8cm}
\end{picture}
\caption{Energy in eV per atom of a free-standing Ga layer as a function
  of the lattice constant. The energy zero refers to a Ga atom. The dashed
  lines mark the lattice constants of GaN and GaAs. Obviously, the GaN lattice
  constant is very close to that of optimum Ga-Ga interaction explaining the
  stabilization of a Ga adlayer found at GaN surfaces.}
\end{figure}

\subsection{Chemical trends} 

The above discussion explains the preference for having exclusively one species
in the surface layer and the tendency to stabilize structures with
low-coordinated configurations that are not observed for conventional III/V
semiconductor surfaces. For these materials, the surface atoms prefer three-fold
coordinated configurations (i.e. surface atoms have a maximum of one dangling
bond) and the dominant surface species changes when going from cation to
anion-rich conditions. Based on the cohesive energies shown in Tab. 1,
we expect similar features (preference of one species, low-coordinated sites) for
the other group III-nitrides (InN, AlN) and also for SiC.

\section{Surface reconstructions on G\protect{a}N surfaces}

The arguments given above are general and apply to any polar GaN surface,
because specific differences between the various surface orientations were not
used. However, as can be seen in Fig. 1 the local configuration and the number
of nearest neighbor bonds at the surface atoms depend strongly on the surface
orientation: for the (001) surface the atoms are two-fold coordinated while for the (111)
and ($\bar1\bar1\bar1$) surfaces they form one or three bonds. We will therefore
elucidate in the following how these specific arrangements affect surface
reconstructions and compare with reconstructions found on GaAs.

\subsection{The cubic GaN (001) surface}

We will start with the cubic (001) surface. For GaAs, a detailed analysis of
STM measurements by Pashley \cite{pashley} revealed a set of rules which
determine the reconstructions of equilibrium surfaces: Surfaces form (2x$N$)
reconstructions where the 2x periodicity arises from the formation of dimers
and the $N$ periodicity arises from the missing surface dimers.  These rules
combined with the electron counting rule largely restrict the number of
possible surface structures. Northrup and Froyen added later the principle of minimizing
the electrostatic energy and identified the (2x4)-$\beta$2 surface (which
obeys all the rules) to have the lowest energy under moderately As-rich conditions
\cite{northrup94}.

In general, dimers are considered to be the natural building block of surface
reconstructions on (001) surfaces and have been observed also for other
materials (e.g. AlAs, Si, SiC). The formation of surface dimers is
energetically favorable since it reduces the number of dangling bond states by
a factor of two. More important, due to the specific arrangement of the atoms
on the (001) surface dimers can be formed simply by rotating the back bonds of
the surface atom without stretching them (see Fig. 5a). While it is apparent
that the formation of dimers is energetically preferred over the
unreconstructed surface, it is by no means obvious why other building blocks
than dimers should not be formed. Looking at the driving forces for dimers
these alternative 'building blocks' should further reduce the number of
dangling bonds without significantly stretching the back bonds of the surface
atoms. The simplest way to do this is to form '$n$-mers' instead of dimers
where $n$ surface atoms form a linear chain (an example for $n$ = 4 is shown
in Fig.  5b). The number of dangling bonds is then reduced by a factor of
2$n$/2 = $n$ where 2$n$ is the number of dangling bond states at the
unreconstructed surface and 2 the number of dangling bond states per
``$n$-mer''. For a dimer we obtain thus a factor of 2 while for a tetramer as
shown in Fig. 5b a factor of 4 is achieved: Using these building blocks the
number of dangling bonds at surfaces can be much more efficiently reduced than
by forming dimers.  The tetramer structure as shown in Fig. 5b has also
another remarkable feature. It obeys electron counting (each of the three bonds in the
tetramer holds 2 electrons and the two remaining dangling bonds
are empty/filled if the tetramer consists of cations/anions). Thus, in contrast
to the dimer structures, which require a combination of dimers and
missing dimers in the surface unit cell, a (1x4) unit cell with a single
tetramer is already sufficient to fulfil the ECR.

We have therefore performed calculations for a Ga tetramer on GaAs and GaN.
For GaAs we find that a tetramer structure is unstable: without any barrier it
spontaneously dissociates into two dimers. The reason for the instability
becomes obvious when looking at the geometry as shown in Fig. 5c. In order to
form the Ga-Ga bonds in the tetramer the back bonds of the outer atoms (marked
by dashed lines) have to be stretched by more than ~20\% implying that these
bonds are virtually broken. For GaAs, the elastic energy necessary to create
the tetramer is larger than the energy gained by reducing the dangling bond
density. For GaN, however, our calculations reveal Ga tetramers to be lower in
energy than a structure consisting of two dimers. In fact, detailed
calculations for a large set of possible surface geometries (including those
stable on GaAs surfaces) revealed that the tetramer structure is actually the
energetically preferred structure \cite{neugebauer98}. From Fig. 5b we see
that the Ga-Ga bonds in the tetramer can be formed almost without stretching
the back bonds of the outer atoms: The length of the back bonds increases only
by ~5\% compared to the unreconstructed surface. The reason is simply given in
terms of geometric ratios. Since the Ga-Ga bond length remains largely
independent of whether the bonds are formed on GaAs or GaN, the stretching of
the back bonds and thus the elastic energy becomes smaller with decreasing
lattice constant (see also Fig. 5b). This explains why tetramers are stable
on GaN ($a_{\rm lat} = 8.49$ bohr) but not on GaAs ($a_{\rm lat} = 10.4$
bohr). Based on these arguments we expect very similar structures for the
other group III-nitrides (AlN, InN).

\setlength{\unitlength}{1mm}
\begin{figure}[tb]\centering
  \psfig{file=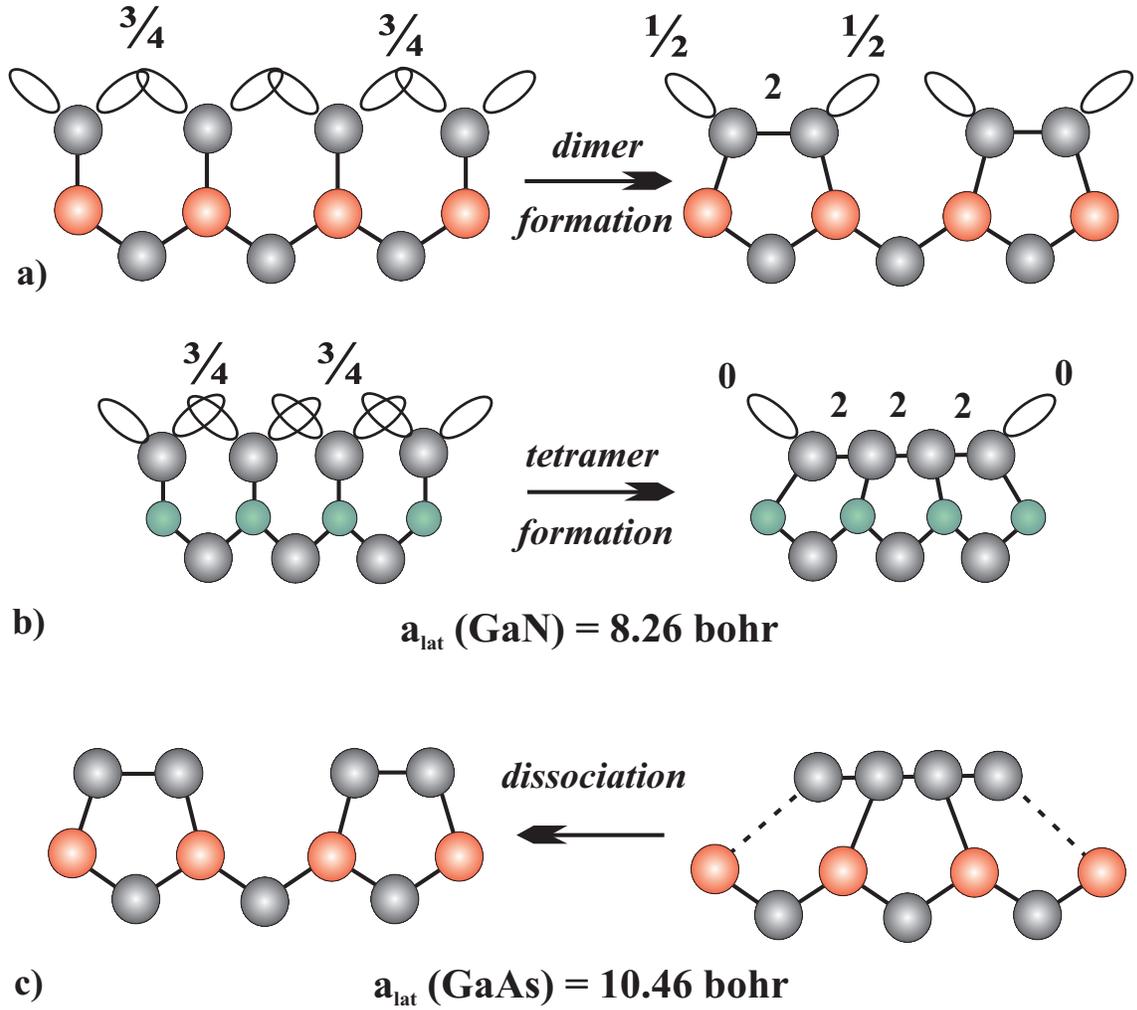,width=15cm}
\caption{Schematic view of the reconstruction mechanisms of the GaN and 
  GaAs (001) surfaces. Note the difference between the lattice constants of
  GaN and GaAs which leads to a completely different reconstruction mechanism
  for the two materials.}
\end{figure}

\subsection{The closed packed (0001) and (000$\bar1$) GaN sufaces}

We will now focus on reconstructions on the close-packed cubic (111) and
($\bar1\bar1\bar1$) structures which are equivalent up to 4th nearest
neighbors with the wurtzite (0001) and (000$\bar1$) surfaces. A main
difference of these surfaces with respect to the (001) surface discussed above
is the hexagonal symmetry ($C_{3v}$) and the lack of a preferred axis along
which dimers can be formed.  Consequently, the formation of dimers always
destroys the point group symmetry of the surface. In fact, dimer geometries
have never been reported for these surface orientations. Experimental and
theoretical studies showed that surface structures are commonly characterized
by the formation of adatom, trimer and vacancy structures \cite{Biegelsen}.
These structures are usually formed in a (2x2) surface unit cell and can be
shown to obey electron counting \cite{Kaxiras}.

Let us now consider the GaN (0001) surface.  As has been pointed out in Sec. 3
the energetically preferred (1x1) structure is the Ga-terminated surface with
one dangling bond state per surface atom (Fig. 1c).  This structure is
compatible with the principles of having mainly Ga-atoms in the surface and a
low dangling bond density.  However, first principles calculations showed that
this structure is not stable: (2x2) adatom structures that satisfy the electron counting
rule are lower in energy \cite{smith,bernholc97}.
These results are also consistent with recent STM
investigations, where mixtures of Ga and N adatoms on these surfaces and
semi-insulating behavior have been found \cite{smithsurf}.

For GaN (000$\bar1$) the tendency to prefer Ga atoms in the surface layer
prevails over all other principles. A Ga-terminated (1x1) structure becomes
energetically most stable despite having a maximum number of dangling bonds
(three per surface atom) and clearly disobeys electron counting (Fig. 1f). Similarly, on the
AlN (000$\bar1$) surface a (1x1) Al-adlayer becomes stable under Al-rich conditions \cite{northrup97}.

\section{Conclusions}

Based on first-principles calculations we have studied the driving forces
governing surface reconstructions on GaN.  The principal mechanisms are (i)
the tendency to stabilize Ga atoms in the surface layer, (ii) to obey electron
counting and (iii) to reduce the number of dangling bonds. While (ii) and
(iii) are well-known mechanisms driving surface reconstructions on other
semiconductor materials (i) is a unique property of GaN, and more generally of all
group-III-nitrides. The strong tendency to stabilize Ga-rich surfaces is not
only a new property without any analogue among other compound semiconductors but
it also prevails over the other principles.  The most extreme example
is the (1x1) Ga-adlayer structure shown in Fig. 1f. This is a metallic structure \cite{smith,feenstra}
in which the number of dangling bonds is maximized.

These rules are of course too simplistic to derive {\it a priori}, i.e. without
input from experiment or without performing realistic calculations.
However, despite their simplicity these
rules have been very successful in explaining or identifying surface
reconstructions for a wide range of semiconductors. The knowledge of these
mechanisms is not only important to derive conclusions concerning surface
reconstructions but also about other properties of the surface. For example,
an immediate consequence of the fact that polar GaN surfaces consists mainly
of Ga atoms is a decreasing surface energy when going from N-rich to Ga-rich
conditions (see Eq. (2)). This might explain why the surface morphology of GaN
appears to improve when growing under more Ga-rich conditions
\cite{schikora,kley}. It also explains the exceptionally low diffusion
barriers of Ga adatoms on these surfaces (0.2 - 0.6 eV) compared to barriers
of 1.5 eV on GaAs \cite{kley}: Since the surface layer is comprised mainly of Ga atoms,
the Ga adatom forms primarily Ga-Ga bonds which are weak compared to the
Ga-N bond (metallic Ga melts already at room temperature). In breaking these
bonds to jump from one adsorption site to the next, the Ga adatom must therefore overcome only
a small energy barrier \cite{zywietz}.

As has been pointed out in detail in Sec. 3 the origin of the stability of
Ga-rich surfaces is (i) the large difference in the energies of the chemical
reservoirs and (ii) a significantly smaller lattice constant compared to
conventional III/V compound semiconductors. We expect therefore a very similar
behavior for the other group III-nitrides, i.e. AlN and InN and their alloys.

\section{Acknowledgements}

We gratefully acknowledge financial support from the BMBF, the ''Fond der
Chemischen Industrie'' (T.Z.), and the ''Deutsche Forschungsgemeinschaft'' (J.N.).

\end{document}